\documentclass[aps,amsmath,amssymb,floatfix,singlecolumn,12pt,prb]{revtex4-1}
\usepackage{graphics}
\usepackage{graphicx}
\usepackage{amssymb}
\usepackage{amsfonts}
\usepackage{amsmath}
\usepackage{bm}
\usepackage{epsf}
\usepackage{version} 
\usepackage{color} 
\usepackage{tikz-cd}

\newcommand \be  {\begin{equation}}
\newcommand \ee  {\end{equation}}
\newcommand \bea {\begin{eqnarray} }
\newcommand \eea {\end{eqnarray}}

\newcommand \bd  {\begin{details}}
\newcommand \ed  {\end{details}}

\begin{document}

\title{Zeeman spin-orbit coupling in antiferromagnetic conductors}
\author{Revaz Ramazashvili} 
\affiliation{Laboratoire de Physique Th\'eorique, Universit\'e de Toulouse, CNRS, UPS, France
}

\date{\today}

\excludeversion{details}


\begin{abstract} 
This article is a brief review of Zeeman spin-orbit coupling, arising in a low-carrier commensurate N\'eel antiferromagnet subject to magnetic field. The field tends to lift the degeneracy of the electron spectrum. However, a hidden symmetry protects double degeneracy of Bloch eigenstates at special momenta in the Brillouin zone. The effective transverse $g$-factor vanishes at such points, thus acquiring a substantial momentum dependence, which turns a textbook Zeeman term into a spin-orbit coupling. 
After describing the symmetry underpinnings of the Zeeman spin-orbit coupling, 
I compare it with its intrinsic counterparts such as Rashba coupling, and then show how Zeeman spin-orbit coupling may survive in the presence of intrinsic spin-orbit coupling. Finally, I outline some of the likely experimental manifestations of Zeeman spin-orbit coupling, and compare it with similar phenomena in other settings such as semiconducting quantum wells.
\end{abstract}

\pacs{75.50.Ee}

\maketitle


\section{Introduction} 
\label{sec:Intro}
 
Spin-orbit coupling, the central character of this Special Issue, appears in non-relativistic quantum mechanics as only a vestige of relativity, in the form of the Pauli term 
$\mathcal{H_P}$ in  the Schr\"odinger Hamiltonian~\cite{Kittel} 
\be
\label{eq:Pauli}
\mathcal{H_P} = \frac{\hbar}{4 m_0^2 c^2} {\bm \sigma} \cdot {\bf p} \times {\bm \nabla} V ({\bf r}),
\ee
where $\hbar$ is the Planck constant, $c$ the speed of light, $m_0$ is the free electron mass, ${\bf p}$ is the electron momentum, ${\bm \sigma}$ its spin, and $V ({\bf r})$ its   potential energy as a function of the electron coordinate ${\bf r}$. In the low-energy expansion, the $\mathcal{H_P}$ arrives as a second-order term in the expansion in weak coupling constant $\alpha = e^2/\hbar c \approx 1/137$. In this sense, spin-orbit coupling is indeed “small”.

Yet in  solids, spin-orbit coupling is responsible for a variety of fundamental phenomena. In magnets, it may induce magnetocrystalline anisotropy, whereby spontaneous magnetization acquires a preferred set of directions with respect to the crystal axes.~\cite{LL-VIII,Cullity} In  antiferromagnets, spin-orbit coupling may give rise to ``weak'' ferromagnetism.~\cite{LL-VIII,Dzialoshinskii-1957,Dzyaloshinsky-1958,Moriya-1960} 
In transition metal compounds, coupling of spin, orbital and structural degrees of freedom leads to a multitude of unusual phases.~\cite{Kugel-1982} In Mott insulators, spin-orbit coupling may produce interesting effects such as realization of an effective Heisenberg-Kitaev model.~\cite{Jackeli-2009} Last but not the least, in an innocuous band insulator spin-orbit coupling may bring to life topologically non-trivial electron states, that have been a subject of much attention.~\cite{Hasan-2010,Moore-2010}

In all of these cases, spin-orbit coupling is {\em intrinsic}: it acts in the absence of any external perturbation applied to the crystal. By contrast, the present article is devoted to the Zeeman spin-orbit coupling, that may appear in a low-carrier N\'eel antiferromagnet, subject to magnetic field. Zeeman spin-orbit coupling is thus a particular example of the Zeeman effect. At the same time, it entangles orbital motion of an electron with its spin, and hence represents a true spin-orbit coupling. 
Zeeman spin-orbit coupling is proportional to the applied magnetic field, and thus is inherently tunable. This distinguishes it from intrinsic spin-orbit coupling, that acts in the absence of any external field, and that most articles in this Topical Issue focus on.

By virtue of the Zeeman spin-orbit coupling, magnetic field splits a single doubly-degenerate electron band into two bands that are non-degenerate almost everywhere in the Brillouin zone. While this is indeed just {\em a} form of the Zeeman effect, there is also a similarity here to how, in the absence of inversion symmetry, intrinsic spin-orbit coupling lifts double degeneracy of an electron band in a non-magnetic crystal. This similarity provides a useful perspective, hence the article opens with Section II, that presents a basic overview of degeneracies in a non-magnetic crystal, with and without external magnetic field. 
Section III does the same for a commensurate collinear N\'eel antiferromagnet, and elucidates the analogy with non-magnetic case. Most of the Section III is a pedagogical presentation of the degeneracies expected in the absence of intrinsic spin-orbit coupling, where the electron spin is entirely decoupled from its orbital motion.~\cite{rr_sym,symlong} The subsection III.c then analyses spectral degeneracies that may appear in transverse magnetic field in the presence of intrinsic spin-orbit coupling. Section IV outlines the implications in the linear order in the field, introduces the Zeeman spin-orbit coupling and recapitulates some of its properties. Section V addresses some of the likely experimental manifestations of the Zeeman spin-orbit coupling, while Section VI offers a brief comparison with other settings, and an outlook.
 
The review covers early studies of the Zeeman spin-orbit coupling in antiferromagnets as well as a case that remains largely unexplored, where antiferromagnetism coexists with substantial intrinsic spin-orbit coupling. I attempted to make the presentation pedagogical and coherent, and hope the reader will find such a review useful. 
 
\section{Spectral symmetries in zero field} 
\label{sec:PM} 

\subsection{A non-magnetic crystal}
\label{subsec:PM-zero-field}

Key insight into spectral degeneracies can be gained by simple symmetry arguments. Consider a non-magnetic crystal, that is one symmetric under time reversal $\theta$. Kramers theorem \cite{Kittel} tells us, that every single-electron Bloch eigenstate $ | {\bf p} \rangle $  at momentum ${\bf p}$ has a degenerate orthogonal partner $\theta | {\bf p} \rangle $ at momentum $- {\bf p}$.~\cite{footnote1} 

\begin{figure}[h]
\center 
\includegraphics[width = 1 \textwidth]{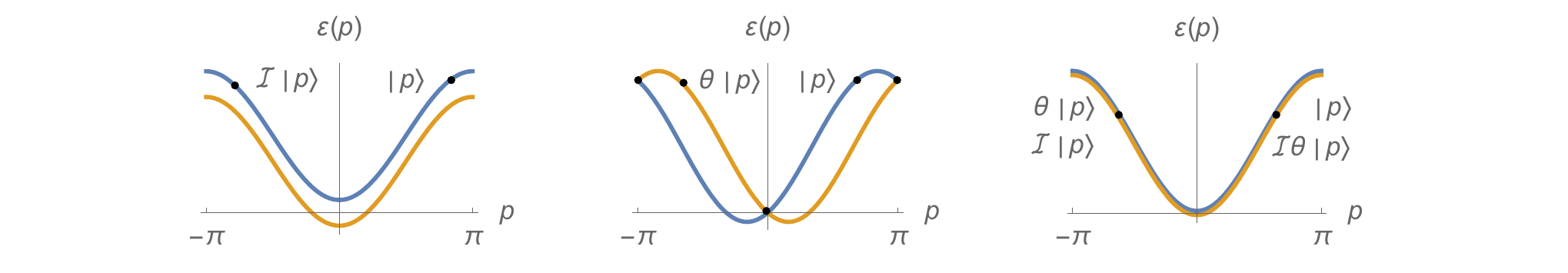}
\caption{(color online). Typical electron spectra $\varepsilon(p)$ in a one-dimensional non-magnetic crystal. {\bf Left panel:} Only the inversion symmetry $\mathcal{I}$ is present, but no time reversal $\theta$. A Bloch eigenstate $| {\bf p} \rangle$ at momentum ${\bf p}$ is degenerate with its partner state $\mathcal{I} | {\bf p} \rangle$ at momentum $- {\bf p}$, 
the two states denoted by black dots. {\bf Central panel:} Only the time reversal symmetry $\theta$ is present, but no inversion symmetry. A Bloch eigenstate $| {\bf p} \rangle$ at momentum ${\bf p}$ is degenerate with its partner state $\theta | {\bf p} \rangle$ at momentum $- {\bf p}$, the two states denoted by black dots. The only two symmetry-protected degeneracies at a given momentum, also marked by black dots, reside at ${\bf p} = 0$ 
and ${\bf p} = \pm \pi$. {\bf Right panel:} Both the $\theta$ and $\mathcal{I}$ are symmetries. In this case, the two degenerate orthogonal states $| {\bf p} \rangle$ and 
$\theta \mathcal{I} | {\bf p} \rangle$ at an arbitrary momentum ${\bf p}$ form a 
quartet with their degenerate partner states $\mathcal{I} | {\bf p} \rangle$ and 
$\theta | {\bf p} \rangle$ at momentum $-{\bf p}$.}
\label{fig:PT}
\end{figure}

But are there degenerate states at a {\em given} momentum? For a generic momentum ${\bf p}$ in the Brillouin zone, time reversal symmetry alone does not protect such a degeneracy -- with an important exception of special momenta ${\bf p}^*$, that are equivalent to their opposite up to a reciprocal lattice vector ${\bf Q}$, so that $-\mathcal{U}{\bf p}^* = {\bf p}^* + {\bf Q}$, where $\mathcal{U}$ is a point symmetry of the lattice. For such a ${\bf p}^*$, the state $\mathcal{U}\theta | {\bf p}^* \rangle$ resides at momentum $-\mathcal{U}{\bf p}^* = {\bf p}^* + {\bf Q}$. The states $\mathcal{U}\theta | {\bf p}^* \rangle$ and $| {\bf p} \rangle$ are degenerate and orthogonal. In the nomenclature of the Brillouin zone, the ${\bf p}^*$ and $- \mathcal{U}{\bf p}^* = {\bf p}^* + {\bf Q}$ are one and the same momentum -- which, therefore, hosts two degenerate orthogonal eigenstates. This can be most simply illustrated in one dimension, where $\mathcal{U}$ is the identity, and ${\bf p}^* = 0$ (the $\Gamma$-point) and ${\bf p}^* = \pm \pi$ are the only such special momenta. At all other momenta, the eigenstates are non-degenerate, as shown in the central panel of the Fig. \ref{fig:PT}.

Attentive reader will instantly recognize that these special momenta ${\bf p}^*$ are nothing but the ``time reversal-invariant momenta'' (TRIM), that have played crucial role in the analysis of topological insulators.~\cite{Fu-2007} However, the role TRIM may play in the appearance of spectral degeneracies was appreciated long before the advent of topological electron systems.~\cite{herring.1937-1,dimmock1,dimmock2}

If, in addition to time reversal, the crystal is symmetric with respect to inversion 
$\mathcal{I}$, then Bloch eigenstates are doubly degenerate in the entire Brillouin zone: at {\em any} momentum ${\bf p}$, the states $| {\bf p} \rangle$ and 
$\mathcal{I} \theta | {\bf p} \rangle$ reside at the very same momentum ${\bf p}$ 
and are orthogonal.~\cite{footnote2}
Put otherwise, if the crystal is symmetric with respect to both the time reversal 
$\theta$ and inversion $\mathcal{I}$, then, by applying $\mathcal{I}$, $\theta$ and 
$\mathcal{I} \theta$ to a Bloch eigenstate $| {\bf p} \rangle$ at an arbitrary momentum ${\bf p}$, one obtains a quartet of degenerate mutually orthogonal states, of which $\mathcal{I} \theta | {\bf p} \rangle$ and $| {\bf p} \rangle$ reside at momentum ${\bf p}$, while $\mathcal{I} | {\bf p} \rangle$ and $\theta | {\bf p} \rangle$ reside at $-{\bf p}$. This is schematically shown in the rightmost panel of the Fig. \ref{fig:PT}.~\cite{footnote.Itheta}

Breaking time reversal symmetry $\theta$ (for instance, by applying magnetic field) lifts the double degeneracy -- generally, at all momenta in the Brillouin zone. However, an arbitrary Bloch eigenstate $| {\bf p} \rangle$ at momentum ${\bf p}$ still has a degenerate orthogonal partner state $\mathcal{I} | {\bf p} \rangle$ at 
momentum $-{\bf p}$, as long as ${\bf p}$ and $-{\bf p}$ are different in the Brillouin zone: 
if ${\bf p} = - {\bf p} + {\bf Q}$, then $ | {\bf p} \rangle$ and $\mathcal{I} | {\bf p} \rangle$ are one and the same state, as illustrated by the leftmost panel of the Fig.~\ref{fig:PT}.
As we saw above, breaking the inversion symmetry while retaining time reversal 
has a similar yet weaker effect: the double degeneracy at a given momentum 
is lifted everywhere {\em except} for those special momenta, that are equal to their opposite in the Brillouin zone. At the same time, a Bloch eigenstate 
$| {\bf p} \rangle$ at an {\em arbitrary} momentum ${\bf p}$ still has a degenerate orthogonal partner $\theta | {\bf p} \rangle$ at momentum $-{\bf p}$.  

These simple arguments show that, in a non-magnetic crystal, it is the inversion 
asymmetry that lifts the double degeneracy of a given electron band almost 
everywhere in the Brillouin zone. The effect can be encapsulated in an intrinsic spin-orbit coupling term $\mathcal{H_{SO}}$ of the form 
\be 
\label{eq:SO}
\mathcal{H_{SO}} = {\bf d}_{\bf p} \cdot {\bm \sigma} ,
\ee  
where ${\bf d}_{\bf p}$ is a real pseudovector, that is an odd function of momentum ${\bf p}$, and ${\bm \sigma}$ is the electron spin. 
Depending on the crystal structure and chemical composition, spin-orbit coupling may take a multitude of forms, whose review would go far beyond the subject of the present article. A concise analysis of symmetry properties of spin-orbit coupling can be found in Refs.  \cite{Rashba.UFN-1965,Samokhin.2009}. An early example of such a coupling was introduced by E. I. Rashba.~\cite{Rashba-1959,Rashba-1960} 

\subsection{Rashba spin-orbit coupling}
\label{subsec:Rashba.SOC}

Rashba spin-orbit coupling \cite{Rashba-1959,Rashba-1960,Bychkov-1984a,Bychkov-1984b} is a form of the 
$\mathcal{H_{SO}}$ of the Eq. (\ref{eq:SO}) in semiconductor quantum wells and heterostructures, where band mismatch at the interface produces electron states that are localized in the transverse direction, but behave as Bloch states in the two dimensions along the interface. For such states, at low momenta, the $\mathcal{H_{SO}}$ in the Eq. (\ref{eq:SO}) reduces to 
\be 
\label{eq:Rashba}
{\mathcal H}_{\mathcal{R}} = \alpha_R ( {\bm \sigma} \cdot {\bf p} \times \hat{\bf n} ) ,
\ee  
where $\alpha_R$ is a material-specific constant, ${\bf p}$ and ${\bm \sigma}$ are the electron momentum and spin operators, and $\hat{\bf n}$ is the normal to the interface.

What may be the order of magnitude of the $\alpha_R$? For a na\"ive estimate, let us turn to the textbook Pauli spin-orbit coupling term of the Eq.~(\ref{eq:Pauli}). If this very term were responsible for the Rashba coupling, what value of $\alpha_R$ would it give rise to? In a quantum well, formed in an elemental material with atomic number $Z$ near a flat interface with vacuum, the average ${\bm \nabla} V_0 ({\bf r})$ points transversely to the boundary, and is of the order of the Coulomb energy $Z e^2/a_B$, divided by the Bohr radius $a_B = \frac{\hbar^2}{Z m e^2}$: $\langle {\bm \nabla} V_0 ({\bf r}) \rangle \sim \frac{Z e^2}{a_B^2} \hat{\bf n}$. Properly averaged~\cite{LL-III}, this yields an estimate $\alpha_R \sim Z^2 \frac{e^2}{\hbar} \left[ \frac{e^2}{\hbar c} \right]^2 \sim v_B \cdot  \left[ \frac{Z}{137} \right]^2$, with $v_B = \frac{e^2}{\hbar}$ being the hydrogen Bohr velocity. Of course, this is not more than a crude order-of-magnitude estimate: a reliable evaluation of spin-orbit coupling parameters such as $\alpha_R$ from first principles is a problem in its own right, as is extracting $\alpha_R$ from experimental data.~\cite{Bychkov-1984a,Bychkov-1984b,Winkler:book,Bindel-2016} In most materials of experimental interest, the ratio $\alpha_R/c$ ($c$ being the speed of light) falls in the range $\alpha_R/c \sim 10^{-4} \div 10^{-2}$. The example of Rashba coupling will prove useful below, as a reference point for the strength of the Zeeman spin-orbit coupling.

\section{Spectral symmetries of a collinear N\'eel antiferromagnet} 
\label{Sec:symmetries.of.Neel}

\subsection{Collinear N\'eel antiferromagnet in zero field}
\label{Subsec:AF-zero-field}
 
Before turning to the Zeeman spin-orbit coupling, let us define the class 
of materials, where it is expected to appear -- a collinear commensurate 
N\'eel antiferromagnet. At each point ${\bf r}$ in the sample, such an antiferromagnet 
is characterized by spontaneous local magnetization density ${\bf M}({\bf r})$, that changes sign upon translation by a period ${\bf a}$ of the crystal lattice: 
${{\bf M}}({\bf r}~+~{\bf a})~=~-~{\bf M}({\bf r})$. The adjective ``collinear'' implies 
that, everywhere in the sample, ${\bf M}({\bf r})$ points along or opposite a single 
direction, one and the same throughout the sample.

In the following, ${\bf M}({\bf r})$ will be treated as static; both thermal and 
quantum fluctuations of magnetic order are thus entirely neglected. This will restrict 
us to the bulk of the antiferromagnetic phase, far from any phase transition, 
thermal or quantum. In particular, this requires temperatures far below both the N\'eel temperature and the electron bandwidth. In practice, this also implies the ordered moment, noticeable on the scale of the Bohr magneton. 
Some of potentially relevant materials were discussed in the Ref. \cite{symlong}.

First, it is helpful to understand the spectral degeneracies of such a magnet in zero field, 
and there is a perfect analogy here with what happens in a non-magnetic crystal, whose spectral symmetries were recapitulated in the Subsection II.a. The local magnetization density ${\bf M}({\bf r})$ couples to the electron spin ${\bm \sigma}$ via the N\'eel exchange coupling $\mathcal{H_N} = {\bm \Delta}_{\bf r} \cdot {\bm \sigma}$, where the ${\bm \Delta}_{\bf r}$ is proportional to ${\bf M}({\bf r})$ and thus inherits its transformation properties: ${\bm \Delta}_{\bf r}$ changes sign upon lattice translation ${\bf T_a}$ as well as upon time reversal $\theta$. Thus, while neither ${\bf T_a}$ nor $\theta$ is a symmetry of the N\'eel state, their product $\theta {\bf T_a}$ is indeed a symmetry.  

In the presence of inversion symmetry, action of $\mathcal{I}$ and of 
$\theta {\bf T_a}$ on an exact Bloch eigenstate $| {\bf p} \rangle$ at an arbitrary momentum ${\bf p}$ generates a quartet of mutually orthogonal degenerate eigenstates: the doublet of $| {\bf p} \rangle$ and $\mathcal{I} \theta {\bf T_a} | {\bf p} \rangle$ 
at momentum ${\bf p}$, and the doublet of $\mathcal{I} | {\bf p} \rangle$ and 
$\theta {\bf T_a} | {\bf p} \rangle$ at $-{\bf p}$, as shown in the left panel of  Fig.~\ref{fig:IthetaT}. Notice the analogy with the quartet of mutually orthogonal degenerate eigenstates in a centrosymmetric non-magnetic crystal, with a doublet of 
$| {\bf p} \rangle$ and $\mathcal{I} \theta | {\bf p} \rangle$ at momentum 
${\bf p}$ and the doublet of $\mathcal{I} | {\bf p} \rangle$ and $\theta | {\bf p} \rangle$ 
at $-{\bf p}$.~\cite{footnote3}

In a non-centrosymmetric antiferromagnetic crystal, there is no single universal 
symmetry to protect a degeneracy of Bloch eigenstates at an arbitrary momentum. 
Moreover, in an antiferromagnet, the Kramers orthogonality relation 
$\langle {\bf p} | \theta | {\bf p} \rangle = - \langle {\bf p} | \theta | {\bf p} \rangle$ 
for a non-magnetic crystal is replaced by a less stringent condition 
\be
\langle {\bf p} | \theta {\bf T_a} | {\bf p} \rangle =
 - e^{- 2 i {\bf p \cdot a}} \langle {\bf p} | \theta {\bf T_a} | {\bf p} \rangle 
 \label{eq:TTa}
 \ee 
 for the scalar product of a given Bloch eigenstate $| {\bf p} \rangle$ at momentum 
 ${\bf p}$ and its symmetry partner $\theta {\bf T_a} | {\bf p} \rangle$ at momentum 
 $- {\bf p}$. The momenta ${\bf p}$ such that ${\bf p} = - {\bf p} + {\bf Q}$ 
 (where ${\bf Q}$ is an {\em antiferromagnetic} reciprocal lattice vector) now 
 fall into two classes.~\cite{Begue-2017,Begue-2018} The first class are those ${\bf p}$, where $e^{- 2 i {\bf p \cdot a}} = 1$ (such as the relevant points at the non-magnetic Brillouin zone boundary, and the point $\Gamma$ at the center of the Brillouin zone); they host a doublet 
 of degenerate states. By contrast, for those ${\bf p}$, where 
 $e^{- 2 i {\bf p \cdot a}} = - 1$, the $\theta {\bf T_a}$ protects no degeneracy. 

Up to the distinction outlined in the preceding paragraph, the analogy between 
a collinear commensurate N\'eel antiferromagnet and a paramagnet is complete. 
In a centrosymmetric material, all bands are doubly degenerate at all momenta 
in the Brillouin zone, be it an antiferromagnet or a non-magnetic crystal. In the 
absence of inversion symmetry, the electron bands are non-degenerate, with the 
exception of special momenta in the Brillouin zone, where the time reversal $\theta$ 
(in a paramagnet) or $\theta {\bf T_a}$ (in an antiferromagnet) protect a double 
degeneracy. Near such a momentum, the two split sub-bands can be described by an effective Hamiltonian with an intrinsic spin-orbit coupling term (\ref{eq:SO}) with ${\bf d}_{\bf p} = 0$ at the degeneracy point. 

Described above, symmetry-protected degeneracies in a commensurate N\'eel antiferromagnet were understood over half a century ago.~\cite{herring1} However, symmetry in the presence of magnetic field was not explored until much later.

\subsection{Collinear centrosymmetric N\'eel antiferromagnet in magnetic field} 
\label{Subsec:Neel-in-field}

A fundamental difference between a non-magnetic crystal and an antiferromagnet 
resides in how the double degeneracy of Bloch eigenstates at a given 
momentum is lifted by magnetic field. In a non-magnetic crystal, the 
Zeeman term $\mathcal{H_Z} = ({\bf H} \cdot {\bm \sigma})$ tends 
to lift the double degeneracy everywhere in the Brillouin zone. In the 
presence of inversion symmetry, this is illustrated in the leftmost panel of 
the Fig. \ref{fig:PT}. 

In an antiferromagnet, the situation is more interesting. Here, the field is applied 
to a system that already has a preferred direction, defined by the staggered exchange field ${\bm \Delta}_{\bf r}$. Thus, the lifting of the double degeneracy by magnetic field ${\bf H}$  may be sensitive to how ${\bf H}$ is oriented relative to ${\bm \Delta}_{\bf r}$. 
It turns out, that a longitudinal field ${\bf H} \| {\bm \Delta}_{\bf r}$ lifts the 
degeneracy everywhere in the Brillouin zone, as illustrated in the leftmost 
panel of the Fig. \ref{fig:PT}. This is hardly surprising. Much more 
interestingly, in a transverse field ${\bf H} \perp {\bm \Delta}_{\bf r}$, the system 
retains enough symmetry to protect double degeneracy of Bloch eigenstates at 
a special set of momenta in the Brillouin zone. 

\begin{figure}[h]
\center 
\includegraphics[width = 0.5\textwidth]{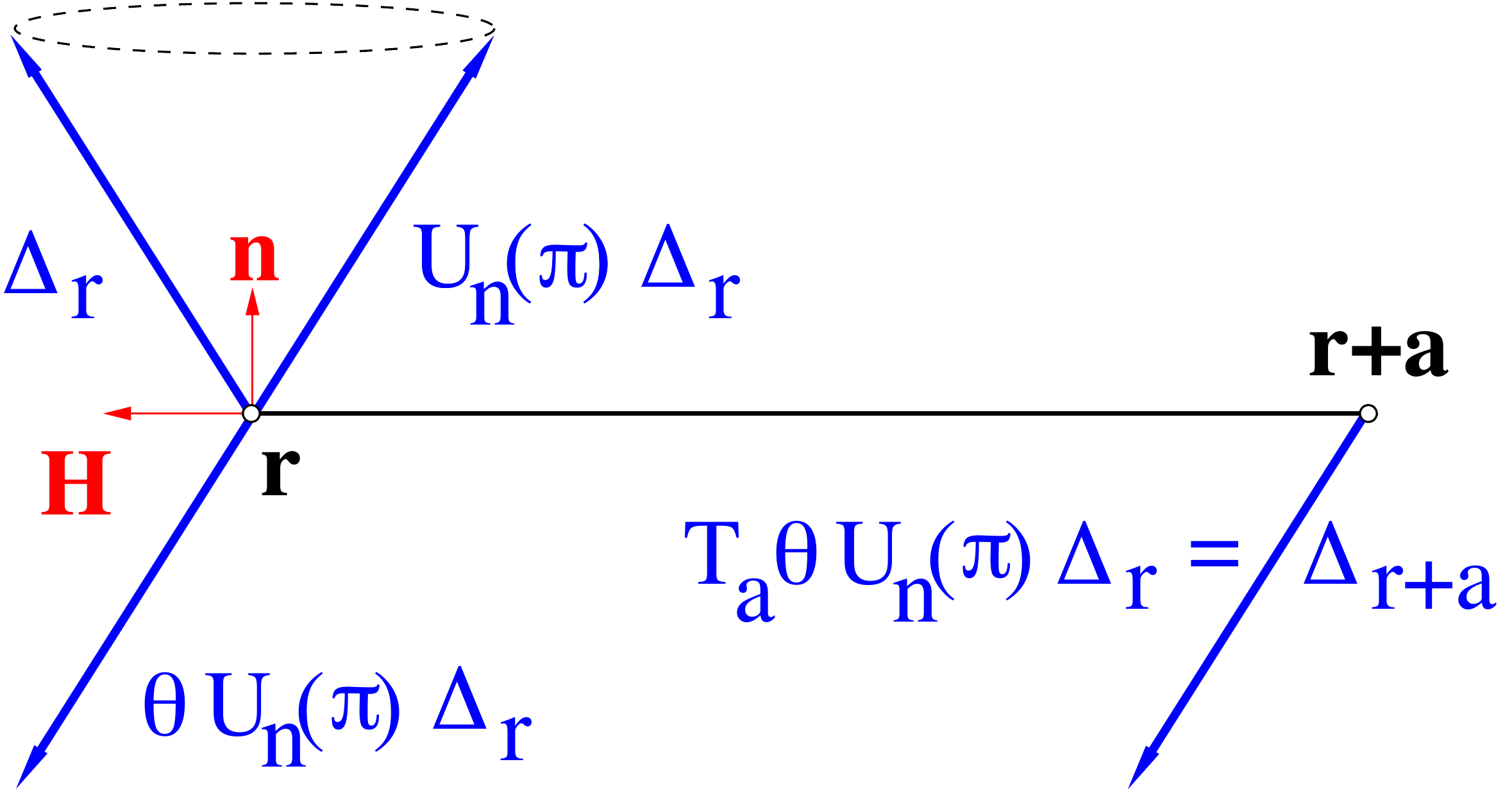} 
\caption{(color online). Two points in space, ${\bf r}$ and ${\bf r} + {\bf a}$, 
separated by half a period ${\bf a}$ of the N\'eel order -- and the exchange 
field ${\bm \Delta}_{\bf r}$ and ${\bm \Delta}_{\bf r + a}$ at these two points, 
upon applying magnetic field ${\bf H}$ transversely to the initial direction 
${\bf n}$ of the staggered magnetization. The figure also shows how the 
exchange field changes upon various transformations, such as 
(i) ${\bf U_n}(\pi)$ -- spin rotation by $\pi$ around ${\bf n}$, 
(ii) $\theta {\bf U_n}(\pi)$ -- combination of ${\bf U_n}(\pi)$ with time reversal $\theta$, 
and (iii) $\theta {\bf U_n}(\pi)$ combined with translation ${\bf T_a}$. To simplify 
notations, the transformations are shown as if they were applied directly to 
${\bm \Delta}_{\bf r}$ rather than to ${\bm \Delta}_{\bf r} \cdot {\bm \sigma}$. 
Notice that, in a finite transverse field ${\bf H}$, the triple product 
$\theta {\bf T_a U_n}(\pi)$ is a symmetry of the tilted N\'eel state.}
\label{fig:canting}
\end{figure}

This symmetry becomes apparent as soon as one pictures a collinear 
centrosymmetric N\'eel state in a transverse field ${\bf H}$, as sketched 
in the Fig. \ref{fig:canting}. 
In a lattice model, the two N\'eel sublattices simply tilt towards ${\bf H}$, 
as do the ${\bm \Delta}_{\bf r}$ in the upper left corner of the figure and 
the ${\bm \Delta}_{\bf r + a}$ in its lower right corner. 
Magnetic moment along the field is invariant upon translation by ${\bf a}$, 
while the moment along the initial direction of the staggered magnetization 
changes sign. More generally, if the physics is not restricted to lattice sites, 
local magnetization is a function of continuous coordinate ${\bf r}$ and, 
in a transverse field, ${\bm \Delta}_{\bf r}$ is no longer collinear. However, 
exactly as in a lattice model, the component ${\bm \Delta}_{\bf r}^\|$ along 
the initial direction of ${\bm \Delta}_{\bf r}$ inherits the N\'eel translation 
antisymmetry: ${\bm \Delta}_{\bf r + a}^\| = - {\bm \Delta}_{\bf r}^\|$. 
By contrast, the component ${\bm \Delta}_{\bf r}^\perp$ along the applied field represents  the field-induced magnetic moment and is translationally symmetric: ${\bm \Delta}_{\bf r + a}^\perp = {\bm \Delta}_{\bf r}^\perp$. 

As we saw above, in zero field the N\'eel exchange term 
$\mathcal{H_N} = {\bm \Delta}_{\bf r} \cdot {\bm \sigma}$ 
was symmetric under $\theta {\bf T_a}$.  Due to collinearity  
of ${\bm \Delta}_{\bf r}$, the $\mathcal{H_N}$ was also 
symmetric under spin rotation ${\bf U_n}(\phi)$ around 
the N\'eel axis ${\bf n}$ by an arbitrary angle $\phi$. 
The ${\bf U_n}(\phi)$ is thus a symmetry of the zero-field Hamiltonian
 -- but {\em only}  in the absence of spin-orbit coupling 
 $\mathcal{H_{SO}} = {\bf d_p} \cdot {\bm \sigma}$, 
 since the latter obviously varies under ${\bf U_n}(\phi)$ 
 due to non-collinearity of the ${\bf d_p}$. 
So, both ${\bf U_n}(\phi)$ and $\theta {\bf T_a}$ are symmetries 
of the longitudinal part ${\bm \Delta}_{\bf r}^\| \cdot {\bm \sigma}$ 
of the exchange term. 
Yet, both of these symmetries are broken by the Zeeman term 
$\mathcal{H_Z} = ({\bf H} \cdot {\bm \sigma})$ and by the field-induced 
term ${\bm \Delta}_{\bf r}^\perp \cdot {\bm \sigma}$, since both change 
sign under $\theta {\bf T_a}$. Remarkably, this can be undone by a 
single uniform rotation ${\bf U_n}(\pi)$, a symmetry of the zero-field 
state.~\cite{footnote4} 
As a result, the combination $\theta {\bf T_a U_n}(\pi)$ 
is a symmetry of the N\'eel state in a finite transverse field. Its action 
on a Bloch eigenstate $| {\bf p} \rangle$ produces a degenerate partner 
eigenstate at momentum $-{\bf p}$. 

Does this symmetry lead to a degeneracy at a given momentum? It does. 
Anti-unitarity of $\theta {\bf T_a U_n}(\pi)$ leads to an analogue of the 
Kramers orthogonality relation~\cite{rr_sym,symlong}
\be
\langle {\bf p} | \theta {\bf T_a U_n}(\pi) | {\bf p} \rangle = 
e^{ - 2 i {\bf p \cdot a}}
\langle {\bf p} | \theta {\bf T_a U_n}(\pi) | {\bf p} \rangle 
\label{eq:TTaUn}
\ee
between the Bloch eigenstate $| {\bf p} \rangle$ at momentum ${\bf p}$ 
and its symmetry partner state at $-{\bf p}$. Those ${\bf p}^*$ that are 
equivalent to their opposite modulo a reciprocal lattice vector ${\bf Q}$ 
of the antiferromagnetic state (${\bf p}^* = - {\bf p}^* + {\bf Q}$), and for 
whom the exponent in the right-hand side of the Eq. (\ref{eq:TTaUn}) 
is different from unity, host a doublet of degenerate states. 

\begin{figure}[h]
\center 
\includegraphics[width = 0.7 \textwidth]{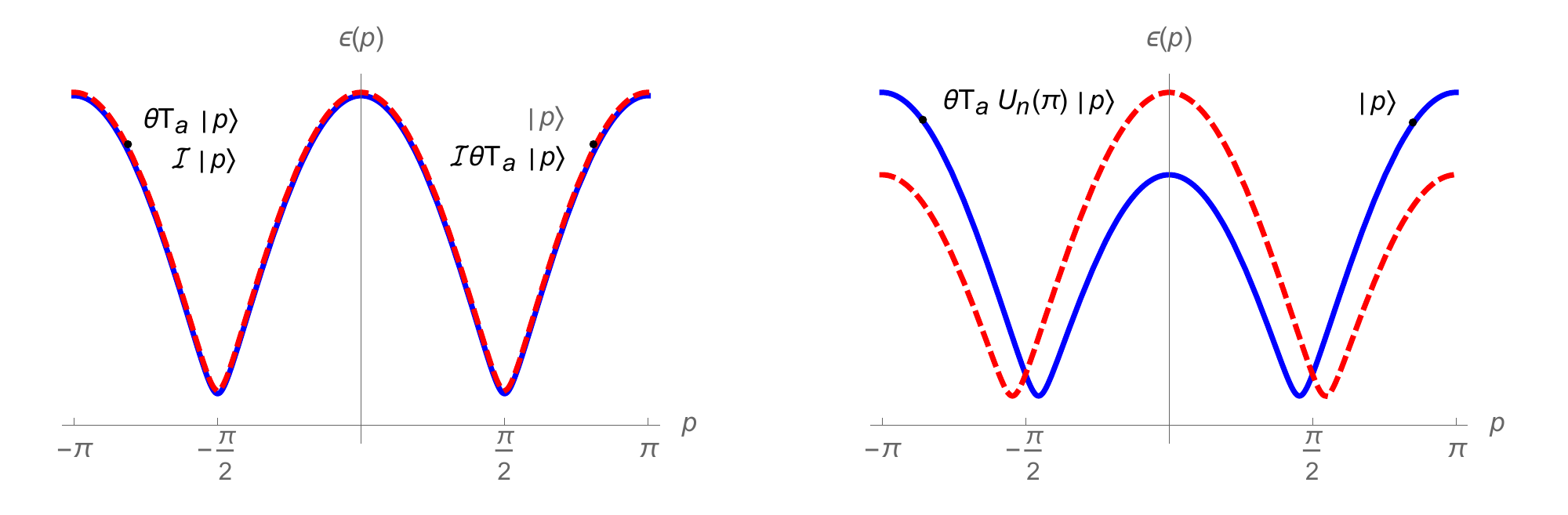}
\caption{(color online). Typical electron spectra $\varepsilon(p)$ in a 
centrosymmetric doubly-commensurate one-dimensional N\'eel antiferromagnet. 
{\bf Left panel:} In zero field, for 
any momentum ${\bf p}$ in the Brillouin zone, the symmetries $\mathcal{I}$ and 
$\theta {\bf T_a}$ generate a quartet of mutually orthogonal degenerate Bloch 
eigenstates: a doublet $| {\bf p} \rangle$ and $\mathcal{I} \theta {\bf T_a} | {\bf p} \rangle$ at momentum ${\bf p}$, and a doublet $\mathcal{I} | {\bf p} \rangle$ 
and $\theta {\bf T_a} | {\bf p} \rangle$ at momentum $-{\bf p}$. 
Each doublet is denoted by black dot.
{\bf Right panel:} In a transverse magnetic field ${\bf H}$, the surviving symmetry 
$\theta {\bf T_a U_n}(\pi)$ protects double degeneracy at the magnetic Brillouin 
zone boundary ${\bf p} = \pm \pi/2$. The degenerate Bloch states $| {\bf p} \rangle$ 
 and $\theta {\bf T_a U_n}(\pi) | {\bf p} \rangle$ are denoted by black dots.}
\label{fig:IthetaT}
\end{figure}

The equation ${\bf p}^* = - {\bf p}^* + {\bf Q}$ implies that such momenta 
${\bf p}^*$ lie at the antiferromagnetic Brillouin zone boundary. However, for those 
${\bf p}^*$ that also belong to the non-magnetic Brillouin zone boundary,~\cite{footnote5} the exponent $e^{ - 2 i {\bf p \cdot a}}$ in the r.h.s. of the Eq. (\ref{eq:TTaUn}) 
equals unity, and thus such a ${\bf p}^*$ does not host a symmetry-protected 
degeneracy. The precise geometry of the set of ${\bf p}^*$ depends on the 
conspiracy between the crystal symmetry and the pe\-rio\-di\-ci\-ty of the 
antiferromagnetic order. A number of possible examples were discussed 
in the Ref. \cite{symlong}. In a one-dimensional doubly-commensurate 
antiferromagnet, the degeneracy is guaranteed at ${\bf p}^* = \pm \pi/2$, 
as illustrated in the Fig. \ref{fig:IthetaT}.

\subsection{N\'eel insulator with intrinsic spin-orbit coupling} 
\label{Subsec:Neel.with.SOC}
 
The arguments above demonstrated how, at a special set of momenta in the Brillouin zone, symmetry protects double degeneracy of the electron spectrum in a centrosymmetric N\'eel antiferromagnet, subject to a {\em finite} transverse magnetic field. However, these arguments tacitly implied both the N\'eel order and the electron spin to be 
decoupled from the underlying crystal lattice. In other words, the intrinsic spin-orbit coupling $\mathcal{H}_{SO}$ of the Eq. (\ref{eq:SO}) was neglected altogether. 

Indeed, as a (formally) small relativistic correction, the $\mathcal{H}_{SO}$ may 
often be ignored. At the same time, the opposite limit of a significant spin-orbit 
coupling is of considerable interest: firstly, intrinsic spin-orbit coupling grows 
rapidly with the atomic number $Z$. In many  antiferromagnets, this alone 
may rule out the possibility of neglecting $\mathcal{H_{SO}}$. Secondly, 
spin-orbit coupling plays a key role in topological properties of condensed matter 
and, recently, a body of work has been devoted to topological properties of 
antiferromagnetic insulators -- see Ref. \cite{Mong-2010} and the subsequent 
work by several groups of authors. 

Topologically non-trivial electron properties are intimately related to degeneracies in 
the electron spectrum. In a N\'eel antiferromagnet {\em without} intrinsic spin-orbit 
coupling, such degeneracies are symmetry-protected even in a finite transverse 
magnetic field~\cite{rr_sym,symlong}. But could symmetry-protected degeneracies 
be present in an antiferromagnet {\em with} substantial intrinsic spin-orbit coupling?~\cite{rr-with-SOC} 

At first sight, this is unlikely. As we saw in Subsections II.a and III.a, 
intrinsic spin-orbit coupling tends to lift the spectral degeneracy everywhere, except for 
a special set of points in the Brillouin zone. It seems that magnetic field could only lift 
any remaining degeneracy. Yet, this is not necessarily the case, as shown below. 

To advance further, we must analyze the symmetries of a Hamiltonian 
that involves, in addition to a common ``non-magnetic'' part, three key terms. 
Firstly, as outlined above, antiferromagnetic order couples to the electron 
spin ${\bm \sigma}$ via the N\'eel exchange coupling 
$\mathcal{H_N} = {\bm \Delta}_{\bf r} \cdot {\bm \sigma}$ 
with collinear field ${\bm \Delta}_{\bf r}$, such that 
${\bm \Delta}_{\bf r + a} = - {\bm \Delta}_{\bf r}$. 
Secondly, an intrinsic spin-orbit coupling of the form 
$\mathcal{H_{SO}} = ( {\bf d_p} \cdot {\bm \sigma})$ involves the field ${\bf d_p}$ 
that is, generally, non-collinear, and transforms as a vector representation of the 
crystal point group. Finally, magnetic field ${\bf H}$ gives rise to the Zeeman term $\mathcal{H_Z} = ({\bf H} \cdot {\bm \sigma})$. 

Without magnetic field, the double degeneracy at a given momentum is protected 
by the very same symmetry $\theta {\bf T_a}$ as in an antiferromagnet without 
intrinsic spin-orbit coupling: the $\mathcal{H_{SO}} = ( {\bf d_p} \cdot {\bm \sigma})$ 
respects both $\theta$ and ${\bf T_a}$ separately, while the N\'eel coupling 
$\mathcal{H_N} = {\bm \Delta}_{\bf r} \cdot {\bm \sigma}$ is symmetric only 
under the product $\theta {\bf T_a}$, which protects the double degeneracy at 
those momenta ${\bf p}$ in the Brillouin zone, that are equivalent to their opposite 
($- {\bf p} = {\bf p} + {\bf Q}$) in the antiferromagnetic Brillouin zone. While 
the set of such momenta in the antiferromagnetic state is different from 
its paramagnetic counterpart, the protecting symmetry $\theta {\bf T_a}$ 
remains the same. 

This simple picture changes once magnetic field is turned on. The reason can be 
traced back to a N\'eel antiferromagnet without intrinsic spin-orbit coupling, where 
Kramers degeneracy in a transverse magnetic field hinges on the combined symmetry 
$\theta {\bf T_a} {\bf U_n} (\pi)$, with ${\bf U_n} (\pi)$ being the spin rotation 
by $\pi$ around the unit vector ${\bf n}$ of the N\'eel magnetization. Generally, 
intrinsic spin-orbit coupling $\mathcal{H_{SO}} = ( {\bf d_p} \cdot {\bm \sigma})$ 
is not invariant under ${\bf U_n} (\pi)$, let alone ${\bf U_n} (\phi)$ with an arbitrary 
rotation angle $\phi$. Thus, with the $\mathcal{H_{SO}}$ present,  
$\theta {\bf T_a} {\bf U_n} (\pi)$ is no longer a symmetry of the problem. 

To begin with, in the presence of $\mathcal{H_{SO}}$ we must specify the orientation 
of the collinear field ${\bm \Delta}_{\bf r}$ with respect to the ${\bf d_p}$; the latter is generally non-collinear and realizes a vector representation of the crystal point group. 
My goal here is not to provide a complete classification, but to demonstrate 
an interesting possibility which is, at the same time, general  enough. Not 
surprisingly, such a possibility appears in a symmetric configuration, where the 
${\bm \Delta}_{\bf r}$ points along a high-symmetry direction of the ${\bf d_p}$. 
For the Bernevig-Hughes-Zhang (BHZ) model \cite{BHZ-2006} with ${\bf d}_{\bf p}^{BHZ} \propto (p_x , p_y , d_z [p_x^2 + p_y^2] )$, the high-symmetry direction is the $z$-axis. Below, I consider the case of the ${\bm \Delta}_{\bf r}$ pointing along the $z$-axis of the ${\bf d}_{\bf p}^{BHZ}$. 
In fact, this choice was implicitly made already in prior work on antiferromagnetic topological insulators,~\cite{Mong-2010} where the ${\bm \Delta}_{\bf r}$ was chosen to point along the symmetry axis $\hat{z}$ of the Bernevig-Hughes-Zhang (BHZ) model \cite{BHZ-2006} in a two-dimensional crystal of square symmetry, with ${\bf d}_{\bf p}^{BHZ} \propto (p_x , p_y , d_z [p_x^2 + p_y^2] )$ near the point $\Gamma$ of the Brillouin zone. 

What is the unitary symmetry of the BHZ spin-orbit coupling 
$\mathcal{H}^{BHZ}_{\mathcal SO} = {\bf d}_{\bf p}^{BHZ} \cdot {\bm \sigma}$? 
Upon spin rotation by angle $\phi$, the ${\bf d}_{\bf p}^{BHZ}$ rotates by the same angle. To make a symmetry, this spin rotation must be compensated by orbital rotation 
$\mathcal{R}_{\bf n} (- \phi)$ by the opposite angle. The $\mathcal{H}^{BHZ}_{\mathcal SO}$ is thus symmetric under the product ${\bf U_n} (\phi) \mathcal{R}_{\bf n} (- \phi)$ with an angle $\phi$, respecting the point symmetry of the ${\bf d}_{\bf p}^{BHZ}$. 

Now, what happens in a field ${\bf H}_\perp$, transverse with respect to ${\bf n}$? 
Without $\mathcal{H}^{BHZ}_{\mathcal SO}$, the product 
$\theta {\bf T_a U_n}(\pi)$ was a symmetry in such a transverse field. 
The $\mathcal{H_{SO}}$ is symmetric under both the $\theta$ and ${\bf T_a}$, 
but not under ${\bf U_n}(\phi)$: as we saw in the previous paragraph, to make 
a symmetry, ${\bf U_n}(\phi)$ had to be combined with the orbital rotation 
${\mathcal R}_{\bf n}(-\phi)$ around the same axis. As a result, a generic Hamiltonian, involving the $\mathcal{H}^{BHZ}_{\mathcal SO}$, the N\'eel exchange 
${\mathcal H}_{\mathcal N} = {\bm \Delta}_{\bf r} \cdot {\bm \sigma}$ and the 
transverse field ${\bf H}_\perp \cdot {\bm \sigma}$, is invariant under the combination 
$\theta {\bf T_a U_n}(\pi){\mathcal R}_{\bf n}(\pi)$. Does this symmetry induce a degeneracy? 

It does, if $\left[ \theta {\bf T_a U_n}(\pi){\mathcal R}_{\bf n}(\pi) \right]^2 \neq 1$. 
Which, in turn, depends on the mutual orientation of ${\bf a}$ and ${\bf n}$. 
If ${\bf a} \perp {\bf n}$, then $\left[ {\bf T_a } {\mathcal R}_{\bf n}(\pi) \right]^2 = 1$, 
hence $\left[ \theta {\bf T_a U_n}(\pi){\mathcal R}_{\bf n}(\pi) \right]^2 = 1$, and the 
symmetry $\theta {\bf T_a U_n}(\pi){\mathcal R}_{\bf n}(\pi)$ induces no degeneracy.~\cite{footnote6} 
In two dimensions, ${\bf a} \perp {\bf n}$ is the only possibility. Thus, in a 
two-dimensional N\'eel antiferromagnet with intrinsic spin-orbit coupling, 
transverse magnetic field lifts the Kramers degeneracy.

\begin{figure}[h]
\center 
\includegraphics[width = 0.3 \textwidth]{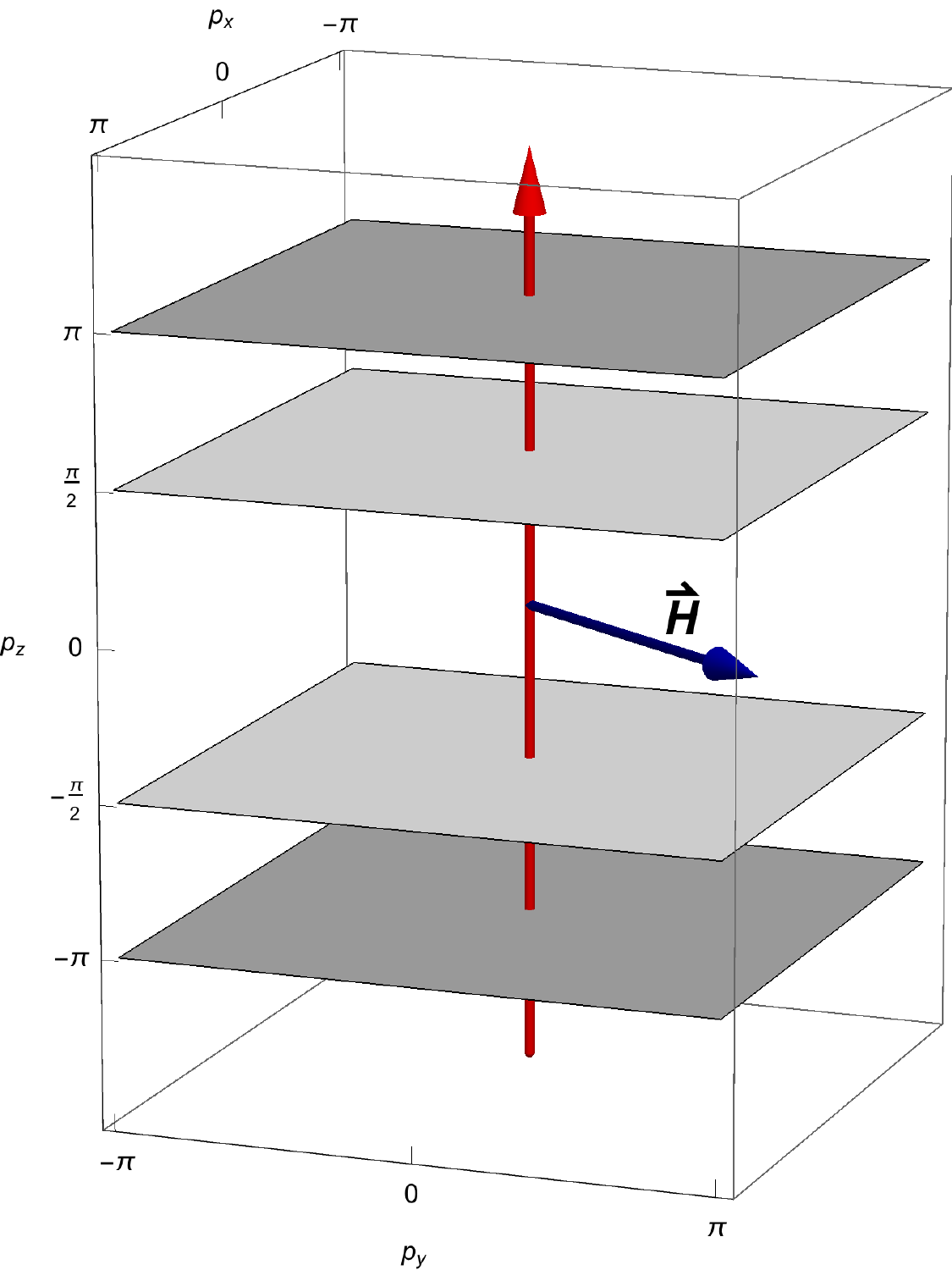} 
\caption{(color online). The Brillouin zone of a three-dimensional N\'eel antiferromagnet with intrinsic spin-orbit coupling. The momentum plane ${\bf p \cdot a} = \pm \pi$ marks the paramagnetic Brillouin zone boundary in the ${\bf z}$ direction, and is shown in darker gray.  As explained in the main text, both the N\'eel half-period ${\bf a}$ and the staggered magnetization ${\bm \Delta}_{\bf r}$ point along the high-symmetry axis ${\bf n}$ of the spin-orbit coupling, that is along the ${\bf z}$ axis of the Brillouin zone, shown by the red arrow. The blue arrow shows the magnetic field, normal to ${\bf z}$. The momentum plane ${\bf p \cdot a} = \pm \pi/2$, shown in lighter gray, marks the antiferromagnetic Brillouin zone boundary in the ${\bf z}$ direction, and hosts doubly degenerate Bloch states.}
\label{fig:AFwithBHZ}
\end{figure}

In three dimensions, the ${\bf a}$ and ${\bf n}$ may be parallel; 
then ${\bf T_a}$ commutes with ${\mathcal R}_{\bf n}(\pi)$, and 
$\left[ {\bf T_a } {\mathcal R}_{\bf n}(\pi) \right]^2 = {\bf T}_{2 {\bf a}}$. 
As a result, when acting on a Bloch eigenstate $| {\bf p} \rangle$,  
$\left[ \theta {\bf T_a U_n}(\pi){\mathcal R}_{\bf n}(\pi) \right]^2 = {\bf T}_{2 {\bf a}} = e^{2 i {\bf p \cdot a}}$. Also, in contrast to the case ${\bf a} \perp {\bf n}$, for ${\bf a} \| {\bf n}$
the state $\theta {\bf T_a U_n}(\pi){\mathcal R}_{\bf n}(\pi) | {\bf p} \rangle$ 
resides at the momentum $-{\bf p}$.~\cite{footnote7} 
In the antiferromagnetic Brillouin zone, the momentum planes ${\bf p \cdot a} = \pm \pi/2$ 
are equivalent, and for such momenta 
$\left[ \theta {\bf T_a U_n}(\pi){\mathcal R}_{\bf n}(\pi) \right]^2 = {\bf T}_{2 {\bf a}}
 = e^{i \pi} = -1$. Thus, at such momenta, the symmetry 
 $\theta {\bf T_a U_n}(\pi){\mathcal R}_{\bf n}(\pi)$ protects double degeneracy 
 in a transverse magnetic field. This is illustrated in the Fig. \ref{fig:AFwithBHZ}.

It is convenient to illustrate the arguments above by the following scheme, 
showing the symmetries of the various terms, and how they are broken as 
new terms are included in the Hamiltonian. The arrows show how the broken 
symmetries combine to form a surviving symmetry. From left to right, the columns 
represent the N\'eel exchange 
${\mathcal H}_{\mathcal N} = {\bf \Delta_r} \cdot {\bm \sigma}$, the intrinsic 
spin-orbit coupling ${\mathcal H}_{\mathcal SO} = {\bf d_p} \cdot {\bm \sigma}$, 
and the combination of the two with the transverse Zeeman term 
${\mathcal H}_{\mathcal Z} = {\bf H}_\perp \cdot {\bm \sigma}$. 

\bigskip 

\begin{tikzcd} 
\label{SO-symmetry-scheme}
{\bf \Delta_r} \cdot {\bm \sigma} & {\bf d_p} \cdot {\bm \sigma} & 
( {\bf d_p} + {\bf \Delta_r} ) \cdot {\bm \sigma} &
( {\bf d_p} + {\bf \Delta_r} + {\bf H}_\perp ) \cdot {\bm \sigma} \\  
 {\bf T}_{2{\bf a}} & {\bf T_a} \arrow[rd] & {\bf T}_{2{\bf a}} &
     {\bf T}_{2{\bf a}}  \\ 
 \theta {\bf T_a} & \theta \arrow[r] & \theta {\bf T_a} \arrow[rd] & \\   
 {\bf U_n}(\phi) \arrow[rr, bend right=20] &  {\bf U_n}(\phi) \mathcal{R}_{\bf n}(-\phi) & {\bf U_n}(\phi) \mathcal{R}_{\bf n}(-\phi) \arrow[r] & \theta {\bf T_a} {\bf U_n}(\pi) \mathcal{R}_{\bf n}(\pi) &  \\ 
\mathcal{R}_{\bf n}(\psi) \arrow[rru, bend right=15] & & &  &  & 
\end{tikzcd}

\section{Zeeman spin-orbit coupling}
\label{Sec:ZSOC}

The arguments of the preceding section have demonstrated a rather peculiar phenomenon: in a commensurate N\'eel antiferromagnet subject to transverse magnetic field, Bloch eigenstates remain doubly degenerate at a set of special momenta in the Brillouin zone. This degeneracy is protected by symmetry even in a finite field -- and, therefore, holds to any order in the field. 

A fundamental consequence of this degeneracy appears already in the first order in the field, that is in the form of the {\em effective} Zeeman term $\mathcal{H}_{\mathcal{Z}}^{eff}$. By its very nature, antiferromagnet has a special direction, set by the staggered magnetization. Of course, other special directions may exist -- defined, for instance, by the crystal structure, but I will first describe the limit where there are none, that is in the absence of any intrinsic spin-orbit coupling. 

In the latter case, the only anisotropy of $\mathcal{H}_{\mathcal{Z}}^{eff}$ is set by the orientation of the magnetic field ${\bf H}$ relative to the unit vector ${\bf n}$ of the staggered magnetization: 
$$
\mathcal{H}_{\mathcal{Z}}^{eff} = - \frac{\mu_B}{2} \left[ 
g_\| ({\bf H}_\| \cdot {\bm \sigma}) + 
g_\perp ({\bf H}_\perp \cdot {\bm \sigma})
\right] . 
$$ 
Here, $\mu_B$ is the Bohr magneton, the ${\bf H}_\| \equiv {\bf n (H \cdot n)}$ 
and ${\bf H}_\perp  \equiv {\bf H} - {\bf H}_\|$ are the longitudinal and transverse 
components of the field with respect to the unit vector ${\bf n}$ of the staggered magnetization, and the $g_\|$ and $g_\perp$ are the longitudinal and transverse 
$g$-factors, respectively. 

Now, in a transverse field, double degeneracy of Bloch eigenstates at certain special momenta ${\bf p} = {\bf p}^*$ in the Brillouin zone means that the $g_\perp$ must vanish at such momenta. Not being identically equal to zero, the $g_\perp$ must, therefore, substantially depend 
on momentum ${\bf p}$, and the $\mathcal{H}_{\mathcal{Z}}^{eff}$ shall be re-written as 
\be
\label{eq:zso}
\mathcal{H}_{\mathcal{Z}}^{eff} = - \frac{\mu_B}{2} \left[ 
g_\| ({\bf H}_\| \cdot {\bm \sigma}) + 
g_\perp ({\bf p}) ({\bf H}_\perp \cdot {\bm \sigma})
\right] . 
\ee
Momentum dependence of the second term in the r.h.s. above, along with the presence of electron spin ${\bm \sigma}$, turns the textbook Zeeman term into a veritable spin-orbit coupling that is the subject of the present article. 

\begin{figure}[h]
\center 
\includegraphics[width = .3 \textwidth]{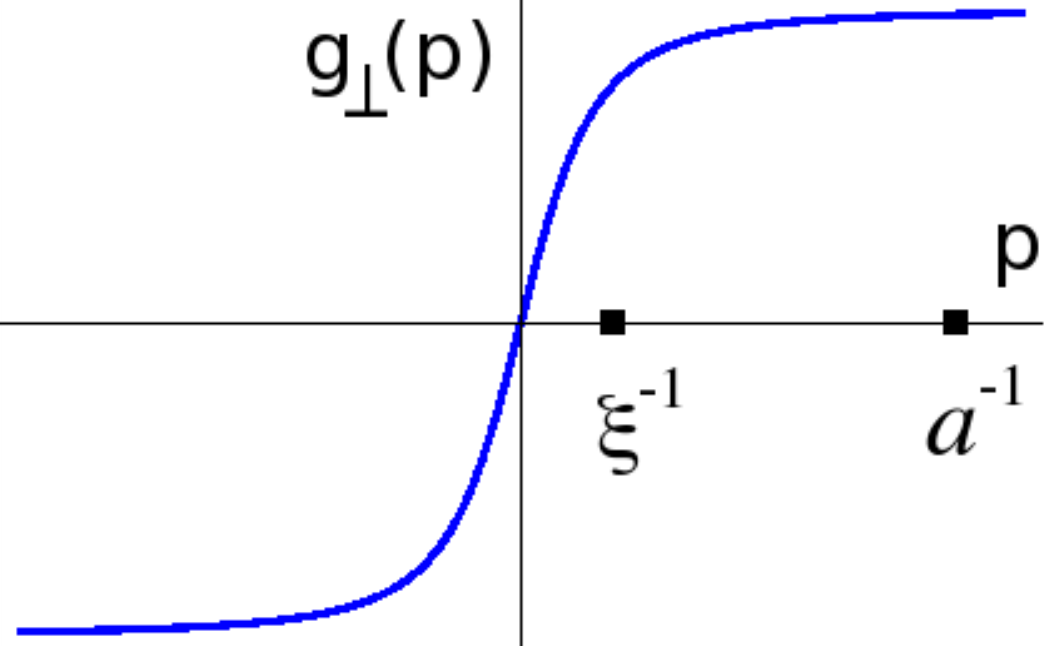}
\caption{(color online). Typical variation of $g_\perp ({\bf p})$ as a function of momentum along the local normal to the degeneracy manifold $g_\perp ({\bf p}) = 0$. Within a momentum range of the order of $\hbar/\xi$, the $g_\perp ({\bf p})$ varies essentially linearly: $g_\perp ({\bf p}) \sim p \xi / \hbar$. Beyond this range, the $g_\perp ({\bf p})$ becomes nearly constant.}
\label{fig:g1D}
\end{figure}

The $g_\|$ is a constant, while general properties of the $g_\perp ({\bf p})$ are as follows.~\cite{symlong} In the absence of intrinsic spin-orbit coupling, the $g_\perp ({\bf p})$ vanishes on a manifold, defined by the equation $g_\perp ({\bf p}) = 0$ in the $d$-dimensional Brillouin zone. This ``degeneracy manifold'' is ($d - 1$)-dimensional. In one dimension, it comprises a set of special points. In two dimensions -- a set of special lines.~\cite{rr-2010,norman-2010} In three dimensions, it forms a set of special surfaces. Nevertheless, over most of the Brillouin zone, the $g_\perp ({\bf p})$ is close to $\pm 1$, and differs from these two values only within a momentum range of the order of $\frac{\hbar}{\xi} \ll \frac{\hbar}{a}$ around the degeneracy manifold. Here $\xi$ is of the order of the antiferromagnetic coherence length \cite{symlong}, and is large on the scale of the lattice spacing $a$. Variation of $g_\perp ({\bf p})$ is thus limited to a momentum range that is small compared with the size of the Brillouin zone. Within a momentum range of about $\frac{\hbar}{\xi}$ around the manifold $g_\perp ({\bf p}) = 0$, the $g_\perp ({\bf p})$ varies linearly with momentum: $g_\perp ({\bf p}) \sim p \xi / \hbar$, where $p$ is measured along the local normal to the degeneracy manifold. Such a variation of $g_\perp ({\bf p})$ is illustrated in the Fig. \ref{fig:g1D}.

This simplest form of the Zeeman spin-orbit coupling (\ref{eq:zso}) is linear in momentum, as is the intrinsic Rashba spin-orbit coupling (\ref{eq:Rashba}). Thus it is instructive to compare the two. 
Firstly, the Zeeman spin-orbit coupling field ${\bf d}_{ZSO}({\bf p}) = - \frac{\mu_B}{2} {\bf H}_\perp \frac{p \xi}{\hbar}$ points along the ${\bf H}_\perp$. By contrast, the spin-orbit field ${\bf d}_R({\bf p}) = \alpha_R \, {\bf p \times \hat{n}}$ of the Rashba coupling is manifestly non-collinear. Secondly, let us compare the two coefficients of the linear momentum dependence. 

For the Rashba coupling, the coefficient is given by the $\alpha_R$. For an elemental material with atomic number $Z$, the na\"ive order-of-magnitude estimate of the Section II.b gave $\alpha_R \sim \frac{e^2}{\hbar} \cdot \left[ \frac{Z}{137} \right]^2$. Relative to the speed of light $c$, this yields $\alpha_R / c \sim \frac{1}{137} \left[ \frac{Z e^2}{\hbar c} \right]^2 \sim \frac{1}{137} \left[ \frac{Z}{137}\right]^2$. In practice, $\alpha_R / c \sim 10^{-4} \div 10^{-2}$. 

By contrast, the Zeeman spin-orbit coefficient relative to $c$ is given by $\alpha_{ZSO}/c \sim \mu_B H_\perp \frac{\xi}{\hbar c} \sim \frac{v_F}{c} \frac{\mu_BH_\perp}{\Delta} \sim \frac{e^2}{\hbar c}\frac{\mu_BH_\perp}{\Delta} \sim \frac{1}{137} \frac{\mu_BH_\perp}{\Delta}$, where the Fermi velocity $v_F$ and the antiferromagnetic coherence length $\xi$ were estimated as per $v_F \sim \frac{e^2}{\hbar}$ and 
$\xi \sim \hbar \Delta / v_F$, with $\Delta$ being the antiferromagnetic gap in the 
electron spectrum. Notice that $\alpha_{ZSO}/c$ carries only a single weak coupling constant $\alpha \approx 1/137$ compared with $\alpha^3$ in the estimate of $\alpha_R / c$.

The ratio of the two couplings is thus $\alpha_{ZSO} / \alpha_R \sim \left[ \frac{\hbar c}{Ze^2} \right]^2 \frac{\mu_B H_\perp}{\Delta} \sim \left[ \frac{137}{Z} \right]^2 \frac{\mu_B H_\perp}{\Delta}$. The present theory applies only in the limit $\frac{\mu_B H_\perp}{\Delta} \ll 1$. However, the factor $\left[ \frac{137}{Z} \right]^2$ is generally large. While this is, indeed, only a hand-waving argument, it indicates that even in a relatively weak field $\frac{\mu_B H_\perp}{\Delta} \ll 1$, Zeeman spin-orbit coupling may be comparable to a rather substantial Rashba coupling. 

\begin{figure}[h]
\center 
\includegraphics[width = .5 \textwidth]{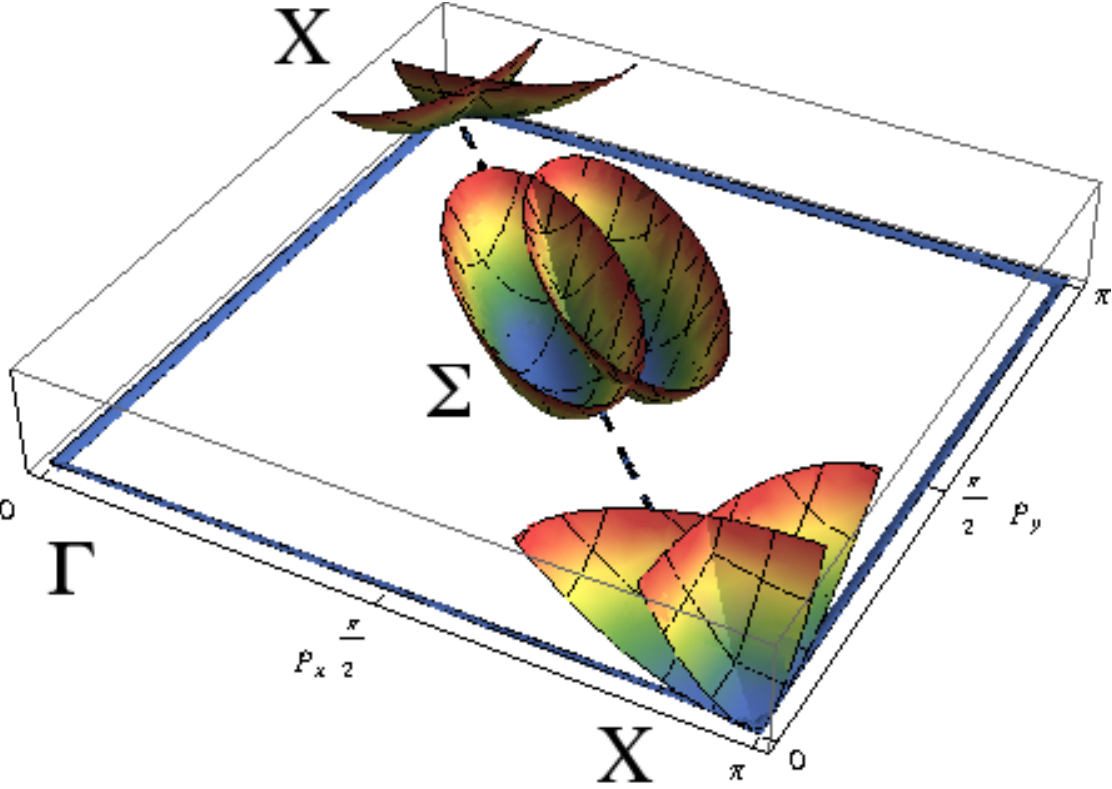}
\caption{(color online). Zeeman splitting of small carrier pockets, notionally centered at the points $X$  and $\Sigma$ in the first quadrant of the Brillouin zone of a two-dimensional N\'eel antiferromagnet on a square lattice. The splitting is induced by magnetic field, transverse to the staggered magnetization. The dashed line, passing through the points $X$ and $\Sigma$, is the magnetic Brillouin zone boundary, where $g_\perp ({\bf p}) = 0$. The pocket sizes and the splitting are exaggerated.}
\label{fig:g_in_BZ}
\end{figure}

Near higher-symmetry points in the Brillouin zone, the degeneracy manifold may cross itself; in this case the $g_\perp ({\bf p})$ shows a more interesting behavior. For example, in a two-dimensional antiferromagnet on a square lattice, the degeneracy lines coincide with the magnetic Brillouin zone boundary $p_x + p_y = \pm \pi$ and $- p_x + p_y = \pm \pi$. Near the corner point $(0, \pi)$ and $(\pi, 0)$ of the magnetic Brillouin zone (point $X$), one finds $g_\perp ({\bf p}) \propto p_x^2 - p_y^2$. At such a point, a degenerate isotropic band extremum $\epsilon({\bf p}) = {\bf p}^2/2m$ is split by transverse field into two ellipsoid sub-bands $\epsilon_\pm ({\bf p}) = \left[ 1 \pm H_\perp/ \Delta \right] p_x^2 / 2m+  \left[ 1 \mp H_\perp/ \Delta \right] p_y^2 / 2m$, where the energy scale $\Delta$ characterizes $g_\perp ({\bf p})$ near point $X$. Such a behavior is shown in the Fig.~\ref{fig:g_in_BZ}.~\cite{rr-2010} The same figure also illustrates linear band splitting near point $\Sigma = (\pi/2 , \pi / 2)$, where a band minimum $\epsilon_\pm ({\bf p}) = p_x^2 / 2m_x + p_y^2 / 2m_y$ is split by transverse field ${\bf H}_\perp$ as per $\epsilon_\pm ({\bf p}) = p_x^2 / 2m_x + p_y^2 / 2m_y - \frac{p_y \xi}{\hbar} ({\bf H}_\perp \cdot {\bm \sigma})$. The degenerate band minimum splits into two identical non-degenerate minima, shifted with respect to each other in momentum space: 
$\epsilon_\pm ({\bf p}) = p_x^2 / 2m_x + \left[ p_y - \frac{m_y \xi}{\hbar} ({\bf H}_\perp \cdot {\bm \sigma}) \right]^2 / 2 m_y$.

What is the leading term in the momentum expansion of the Zeeman spin-orbit coupling $g_\perp ({\bf p}) ({\bf H_\perp \cdot \bm{\sigma}})$ around the degeneracy planes $p_z = \pm \pi/2$, in the presence of an intrinsic spin-orbit coupling, described in the Subsection III.c? 
The symmetry of $g_\perp ({\bf p}) ({\bf H_\perp \cdot \bm{\sigma}})$ must match that of the Hamiltonian. In particular, it must be invariant under $\theta {\bf T_a} {\bf U_n}(\pi) \mathcal{R}_{\bf n}(\pi)$. This implies $g_\perp (p_x , p_y , p_z) = g^*_\perp (p_x , p_y , - p_z)$, while the degeneracy means $g_\perp (p_x , p_y , \pm \pi/2) = 0$. Since $g_\perp ({\bf p}) ({\bf H_\perp \cdot \bm{\sigma}})$ is Hermitian, a linear term in $p_z$ with an imaginary coefficient is forbidden. The leading term allowed is thus $\mathcal{H}_{\mathcal{Z}}^{eff} \propto p_z^2 ({\bf H_\perp \cdot \bm{\sigma}})$. 

\section{Experimental manifestations} 
\label{Sec:experiment} 

Zeeman spin-orbit coupling may manifest itself in a number of ways, that all stem from field-induced entanglement of the electron spin with its orbital motion. As a result, the Landau level spectrum and its Zeeman splitting acquire an unusual dependence on the field orientation \cite{kabanov,rr,zedr} that, in turn, has a number of experimental consequences.

\subsection{Magnetic quantum oscillations as a diagnostic tool}
\label{Subseq:MQOS}

One such consequence is that, in a purely transverse field, Landau levels undergo no Zeeman splitting \cite{kabanov,rr,zedr} -- if the carrier pocket is centered at a symmetry-protected degeneracy point such as those shown in the Fig.~\ref{fig:g_in_BZ}.~\cite{rr-2010,norman-2010} In magnetic quantum oscillations, Zeeman splitting of the Landau levels produces the so-called `spin zeros' \cite{schoenberg}, that is special field orientations, where the oscillation amplitude vanishes. In an antiferromagnetic insulator, absence of spin zeros may thus constrain the precise position of the band extremum in the Brillouin zone, or even pinpoint it exactly.~\cite{rr-2010,norman-2010} This could be particularly useful in antiferromagnets of complex structure and chemical composition, where the location of the band extrema in the Brillouin zone may be less than obvious.

\subsection{Zeeman electric-dipole resonance} 
\label{Subsec:ZEDR}

In addition to thermodynamic measurements such as magnetic quantum oscillations, the Landau level spectrum can be studied by resonant spectroscopy. According to textbook, AC electric field excites spin-conserving cyclotron resonance (CR) transitions between adjacent Landau levels, at the Larmor frequency, whereas AC magnetic field excites spin flip (spin resonance, ESR) transitions at the Zeeman frequency, at a fixed Landau level. Intrinsic spin-orbit coupling enriches this simple picture rather dramatically: it allows one to induce ESR transitions by an AC {\em electric} rather than magnetic field -- and, more generally, to induce transitions at combined frequencies. Predicted by E. I. Rashba over fifty years ago,~\cite{Rashba.UFN-1965} such transitions are called ``electric-dipole spin resonance'' or ``combined resonance". In case of Zeeman spin-orbit coupling, such transitions may be called Zeeman electric-dipole resonance; their theory was developed in the Ref. \cite{zedr}.

The term `electric-dipole resonance' implies that the resonance arises from a dipole term $e {\bf E \cdot r}$ in the Hamiltonian, where ${\bf E}$ is the external electric field, and ${\bf r}$ is the displacement. The resonance matrix elements are thus determined by the characteristic scale of ${\bf r}$. For the textbook ESR, this length scale is given by the Compton length $\lambda_C = \frac{\hbar}{mc} \approx 0.4$ pm. By contrast, for the Zeeman electric-dipole resonance, the length scale in question is the antiferromagnetic coherence length $\xi = \hbar v_F/\Delta$,~\cite{zedr} where $\Delta$ is the antiferromagnetic gap in the electron spectrum. 

Thus, matrix elements of Zeeman electric-dipole resonance exceed those of ESR by about 
$\frac{\hbar c}{e^2} \cdot \frac{\epsilon_F}{\Delta} \approx 137 \cdot \frac{\epsilon_F}{\Delta}$, or at least by two orders of magnitude. Resonance absorption is proportional to the square of the transition matrix element; thus the absorption due to electric excitation of spin transitions exceeds that of ESR at least by four orders of magnitude.

Last but not the least, Zeeman electric-dipole resonance absorption shows a non-trivial dependence on the orientation of the AC electric field with respect to the crystal axes, and on the orientation of the DC magnetic field with respect to the staggered magnetization. 
  
\section{Summary and outlook}
\label{Sec:Summary}

As argued above, Zeeman spin-orbit coupling in a N\'eel antiferromagnet is induced by magnetic field and arises due to a hidden symmetry, that protects double degeneracy of electron eigenstates at special momenta in the Brillouin zone. These special momenta form a degeneracy manifold, whose dimensionality is reduced against that of the Brillouin zone. Limited to special momenta, the degeneracy means that the transverse $g$-factor acquires a substantial momentum dependence which is, however, limited to a relatively small part of the Brillouin zone. Therefore, observation of Zeeman spin-orbit coupling requires that carriers be limited to this small part of the Brillouin zone. Which, in turn, suggests low-carrier antiferromagnetic conductors as a likely system with Zeeman spin-orbit coupling. An extended discussion of relevant materials and experimental constraints may be found in the Refs.~\cite{symlong,zedr}. 

\subsection{Zeeman spin-orbit coupling in semiconducting quantum wells} 
\label{Subsec:wells}

To gain a better perspective, it is instructive to look for Zeeman spin-orbit coupling in materials other than antiferromagnets. A welcome example is provided by III-V direct band gap semiconductors of zinc-blende structure, such as GaAs and InAs. Before turning to details, let us recall that, in the simplest case of well-separated conduction and valence bands, it is the intrinsic spin-orbit coupling that may render the $g$-tensor momentum-dependent. In a centrosymmetric system, spin-orbit coupling has no intraband matrix elements \cite{Rashba.UFN-1965,Samokhin.2009}, and thus the corrections it induces to the $g$-tensor are small at least in the measure of $\Delta_{SO}/E_0 \ll 1$, where $\Delta_{SO}$ is the interband matrix element of the intrinsic spin-orbit coupling, and $E_0$ is the band gap.~\cite{Roth.1959} The same small ratio limits the relative variation of the $g$-tensor across the Brillouin zone.

\begin{figure}[h]
\center 
\includegraphics[width = .4 \textwidth]{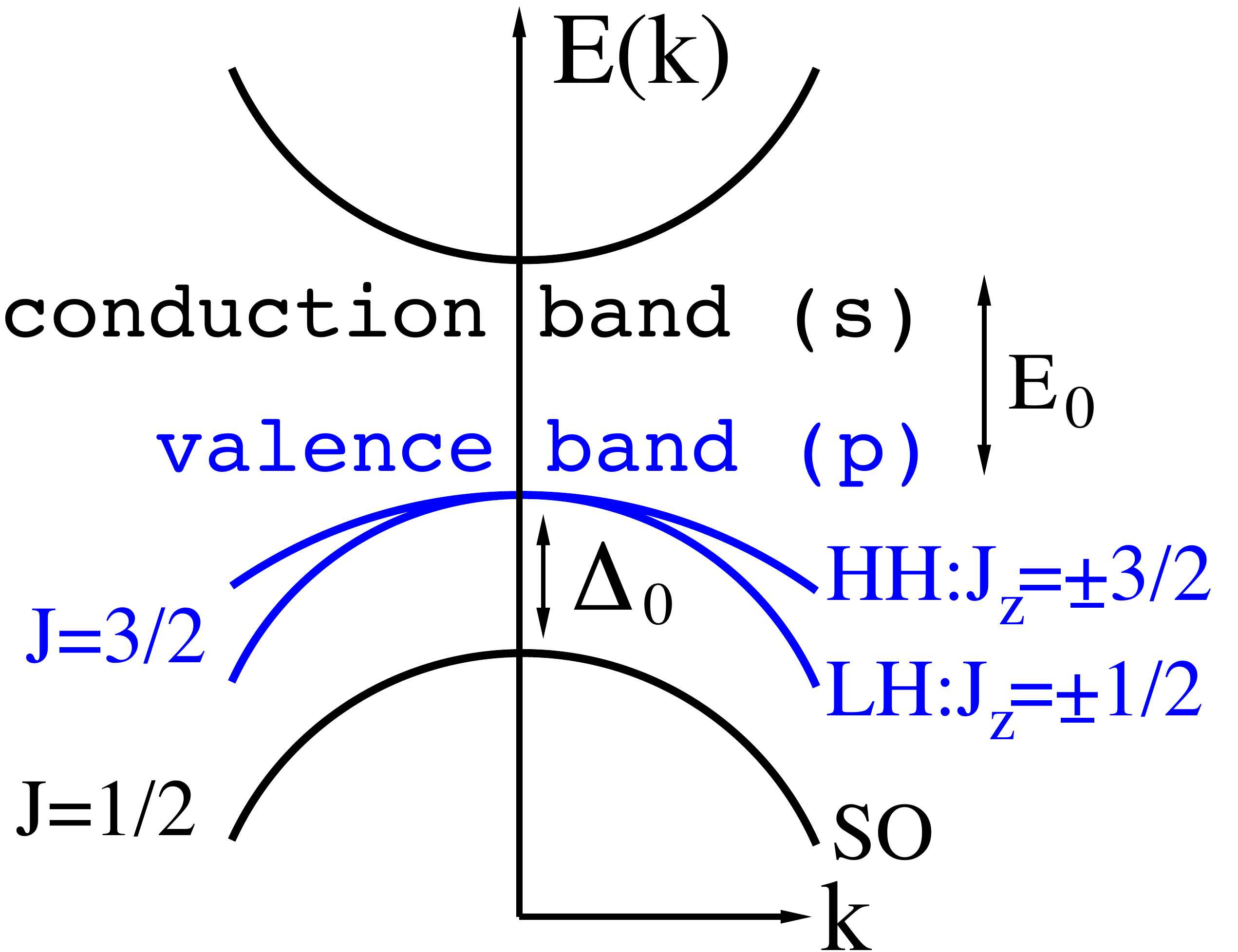}
\caption{(color online). Band structure of GaAs, sketched near fundamental gap $E_0$ in the vicinity of the $\Gamma$-point. As explained in the main text, the conduction band is formed by $s$-orbitals while the valence band is formed by $p$-orbitals. The valence band  ($J = 3/2$) splits into $J_z = \pm 3/2$ heavy hole (HH) band and $J_z = \pm 1/2$ light hole (LH) band, touching at the $\Gamma$-point. The HH and LH bands are separated from the $J = 1/2$ band by the spin-orbit gap $\Delta_0$.
}
\label{fig:GaAs}
\end{figure}

This picture becomes more involved for touching bands, and also if other bands are present nearby, which is the case at the $\Gamma$-point of bulk GaAs and InAs (see Fig. \ref{fig:GaAs}). Here, the conduction band is made mostly of $s$-orbital states, whereas the holes are of $p$-orbital nature. At the $\Gamma$-point, the six hole $p$-states split into a $J = 3/2$ quartet and a $J = 1/2$ doublet, separated from the quartet by the `spin-orbit gap' $\Delta_0$. Upon leaving the $\Gamma$-point, the quartet splits into two doubly degenerate bands: the predominantly $J_z = \pm 3/2$ `heavy hole' (HH) band and its predominantly $J_z = \pm 1/2$ `light hole' (LH) counterpart. Yet, in bulk GaAs, the electron $g$-factor remains isotropic, with $g \approx -0.44$, whereas in bulk InAs  $g \approx -15$; in both materials relative variation of $g$ across the Brillouin zone is of the order of $0.04$.~\cite{Ivchenko:book}

Effective anisotropy can be enhanced by crossing over from a bulk sample to nearly two-dimensional semiconducting quantum well. Due to a finite size of the well, hole motion in the growth direction becomes quantized, and interplay of the reduced dimensionality with the $p$-state anisotropy of holes brings about many interesting effects, that have been a subject of active ongoing study.~\cite{Winkler:book,Ivchenko:book,Dyakonov:book} Here, we will be interested only in the properties of the hole $g$-tensor and, for simplicity, will consider only  field along the plane of the quantum well. Level separation due to finite size of the well in the growth direction will be assumed to greatly exceed all the other relevant energy scales such as the spin-orbit gap $\Delta_0$ and the hole cyclotron frequencies.  

Zeeman splitting of both the `heavy holes' HH ($J_z = \pm 3/2$ at momentum ${\bf k} = 0$) and `light holes' LH ($J_z = \pm 1/2$ at ${\bf k} = 0$) in  magnetic field ${\bf H}$ arises~\cite{Winkler:book} from the following two terms: 
\be
\label{eq:Luttinger-Zeeman}
\mathcal{H} = - 2 \mu_B \left[ \kappa ({\bf J \cdot H}) + q (J_x^3 H_x + J_y^3 H_y + J_z^3 H_z) \right] ,
\ee
where $\mu_B$ is the Bohr magneton, vector ${\bf J} = (J_x , J_y , J_z )$ is made of the $4 \times 4$ matrices of the angular momentum $J = 3/2$, and $\kappa$ and $q$ are parameters of the Luttinger Hamiltonian.~\cite{Luttinger-1956} Notice that the first term above is spherically symmetric, whereas the second term arises because the bulk symmetry is cubic rather than spherical. Accordingly, the coefficient $\kappa$ is of the order of unity, whereas $q$ tends to be small: in GaAs,  $\kappa \approx 1.2$ and $q \approx 0.04$.~\cite{Ivchenko:book} For light holes at ${\bf k} = 0$, the first term in the Eq.~(\ref{eq:Luttinger-Zeeman}) produces non-zero splitting proportional to $\kappa$; the light hole $g$-factor is thus of the order of unity.~\cite{Kiselev-2001} By contrast, for heavy holes at the $\Gamma$-point $\delta J_z = 3$, and the first term of the Eq.~(\ref{eq:Luttinger-Zeeman}) does not contribute to the Zeeman effect. Zeeman splitting of heavy holes arises from the second term in the Eq.~(\ref{eq:Luttinger-Zeeman}) and is small in the measure of $q \approx 0.04$.~\cite{Kiselev-2001} At non-zero ${\bf k}$, admixture of $J_z = \pm 1/2$ components to the heavy holes produces contributions to the $g$ factor, proportional to $k^2$, $k^4$ and so on: 
\be
\label{eq:gk}
g({\bf k}) = g_0 + (ka)^2 g_2 (\phi) + (ka)^4 g_4 (\phi) + ... , 
\ee
where $a$ is mostly defined by the characteristic width of the well in the growth direction, $g_0 \propto q$ is a constant, and $g_2 (\phi)$ and $g_4 (\phi)$ are functions of the momentum direction in the plane of the well.~\cite{Marie-1999,Winkler-2000,Winkler:book,Miserev-PRB-2017,Miserev-PRL-2017}  Hole carriers in a GaAs quantum well are thus subject to Zeeman spin-orbit coupling. Let us compare the latter with its counterpart in a N\'eel antiferromagnet:

Firstly, in semiconducting quantum wells Zeeman spin-orbit coupling appears  due to  anisotropy, enhanced by reduced dimensionality of the well. By contrast, in a N\'eel antiferromagnet, Zeeman spin-orbit coupling emerges due to conspiracy of the anti-unitary symmetry of the N\'eel order with the symmetry of the crystal lattice. As shown in Section III.c, in an antiferromagnet intrinsic spin-orbit coupling and reduced dimensionality may play a subsidiary role.

Secondly, in a quantum well Zeeman spin-orbit coupling appears on the background of a non-zero constant $g$-factor: one may divide the in-plane components of the $g$-tensor into a momentum-independent term $g_0 \neq 0$ and a momentum-dependent part $\delta g({\bf k})$, so that $g({\bf k}) = g_0 + \delta g({\bf k})$, with $\delta g({\bf 0}) = 0$. The functional form of $\delta g({\bf k})$ and its scale relative to $g_0$ depend on the width of the well and its growth direction. In GaAs quantum wells, some studies have found $\delta g({\bf k}) \ll g_0$,~\cite{Marie-1999}, while others~\cite{Winkler:book,Winkler-2000,Gradl-2018} found the scales of $\delta g({\bf k})$ and $g_0$ to be comparable for most growth directions -- and numerically small, with typical measured values of $g$ in the plane being of the order of 
$10^{-2}$.~\cite{Marie-1999,Winkler-2000,Winkler:book,Gradl-2018} Notionally reducing $\delta g({\bf k})$ at a constant $g_0$ (for instance, by increasing the width of the quantum well), this picture can be continuously tuned to the textbook case of the constant momentum-independent $g$-tensor, and Zeeman spin-orbit coupling can thus be ``switched off''. In a N\'eel antiferromagnet, symmetry-protected zeros of the transverse $g$-factor are qualitatively different: here $g_0 = 0$ while the characteristic scale of $\delta g({\bf k})$ is unity, and Zeeman spin-orbit coupling can be ``switched off'' only by destroying the N\'eel state.

Quantum wells grown in the high-symmetry direction [111] are an exception to the above. Such wells are symmetric under $2\pi/3$ rotation around the growth axis, whereas the second term in the Eq.~(\ref{eq:Luttinger-Zeeman}) is not, and thus its Zeeman matrix elements vanish: in such a well, $g_0 = 0$ is protected by the symmetry of growth direction, just as in an antiferromagnet $g_0 = 0$ is protected by a hidden anti-unitary symmetry.~\cite{footnote:[001]} However, in contrast to an antiferromagnet, the momentum-dependent part $\delta g({\bf k})$ in GaAs is still small compared with unity. Much greater bulk $g$-factor ($g \approx 15$) makes InAs quantum wells promising in this regard.~\cite{Matsuo-2017}

Thirdly, III-V semiconductors of zinc-blende structure lack inversion center. Therefore, in a generic quantum well, both HH and LH bands are split by Dresselhaus and Rashba spin-orbit couplings.~\cite{Rashba-1988,Durnev-2014} This splitting may be significant, thus obscuring the Zeeman effect in general and Zeeman spin-orbit coupling in particular. By contrast, in a centrosymmetric N\'eel antiferromagnet such effects are absent. Moreover, as we saw in Section III.c, in an antiferromagnet Zeeman spin-orbit coupling may be symmetry-protected even in the presence of intrinsic spin-orbit coupling.

To summarize, both weakly-doped antiferromagnetic insulators and III-V semiconducting quantum wells have their stronger and weaker points for the study of  Zeeman spin-orbit coupling. GaAs  quantum wells have been studied in great detail. At the same time, typical values of in-plane hole $g$-factor in this material are small. Semiconductors with greater $g$-factor, such as InAs, hold promise in this regard. By contrast, in a low-carrier N\'eel antiferromagnet, Zeeman spin-orbit coupling is protected by symmetry. At the same time, many such materials have not yet been as well-characterized as III-V semiconductors. Clearly, more studies are called for. 

\subsection{Other possibilities}
\label{Subsec:Other}

One may ask whether analogues of the Zeeman spin-orbit  coupling may arise in a different physical context. Indeed, one such possibility is known in the cold-atom physics, where synthetic spin-orbit coupling may be generated and tuned in an experiment, involving laser-induced Raman transitions between two internal states of an atom playing the role of its `pseudospin' (for a review, see \cite{galitski-2013} and references therein). Synthetic spin-orbit coupling can be presented as an artificial gauge field. Just as Zeeman spin-orbit coupling, synthetic spin-orbit coupling in cold-atom setups is tunable, which holds great promise in spite of numerous experimental challenges \cite{galitski-2013}. 

Another closely related and very interesting development involves non-symmorphic magnetic crystals~\cite{fang-2015,brzezicki-2017}, that may give rise to degeneracies akin to those described above -- and, therefore, to some of the similar effects. So far explored very little, this direction may bring some positive surprises. 

\subsection{Acknowledgements}

It is my pleasure to thank Prof. P. Pujol for inspiring conversations on the subject, and Profs.  M. M. Glazov, Y. B. Lyanda-Geller and E. Ya. Sherman for helpful discussions of the physics of quantum wells. 

\section{Appendix: orthogonality relation and Kramers theorem}

This Appendix proves the relation 
\be
\label{eq:antiproduct}
\langle
\phi |
\left[ \mathcal{O} \theta \right]^+ 
     | 
 \left[
 \mathcal{O} \theta 
 \right]
 | 
\psi
\rangle
 = 
\langle 
\psi 
 | 
\phi
\rangle  
\ee 
and points out some of its consequences. Here $| \phi \rangle$ and $| \psi \rangle$ 
are arbitrary states,  $\mathcal{O}$ 
is an arbitrary unitary operator, and $\theta$ 
is time reversal. In the main text, this relation 
is used for 
$| \phi \rangle = \mathcal{O} \theta | \psi \rangle$; 
in this case, when read right to left, 
Eqn. (\ref{eq:antiproduct}) yields 
\be
\label{eq:antiproduct2}
\langle 
\psi 
 |
\mathcal{O} \theta | \psi \rangle 
 =
\langle
\psi 
 |
[ \left(\mathcal{O} \theta \right)^+ ]^2
 | 
 (
 \mathcal{O} \theta )
 |
\psi
\rangle.
\ee  
Whenever $| \psi \rangle$ is an eigenvector of the linear operator 
$[ \mathcal{O} \theta ]^2$ with an eigenvalue different from unity, the 
Eqn. (\ref{eq:antiproduct2}) proves orthogonality of $| \psi \rangle$ and 
$ \mathcal{O} \theta | \psi \rangle$. 

The proof of Eqn. (\ref{eq:antiproduct}) is based 
on the obvious relation 
$(\mathcal{C} \phi, \mathcal{C} \psi)
 = (\psi, \phi)$ for arbitrary complex vectors 
$\phi$ and $\psi$, where 
$(\psi, \phi) \equiv \sum_i \psi_i^* \phi_i$ 
denotes scalar product, and $\mathcal{C}$ 
is complex conjugation. 
Hence, for an arbitrary unitary operator 
$\mathcal{O}$, one finds 
$(\mathcal{O} \mathcal{C} \phi, \mathcal{O} \mathcal{C} \psi) 
 = (\psi, \phi)$, due to invariance of scalar product 
under unitary transformation. Time reversal $\theta$ 
can be presented as a product of $\mathcal{C}$ and a 
unitary operator \cite{wigner}: 
$\theta = \mathcal{V} \mathcal{C}$ 
, thus 
$\mathcal{C} = \mathcal{V}^{-1} \theta$ and, therefore, 
$(\mathcal{O} \theta \phi, \mathcal{O} \theta \psi) 
 = (\psi, \phi)$. 
As a result, for arbitrary states 
$| \psi \rangle$ and $| \phi \rangle$,  one finds 
$
\langle
\phi |
\left[ \mathcal{O} \theta \right]^+ 
     | 
 \left[
 \mathcal{O} \theta 
 \right]
 | 
\psi
\rangle
 = 
\langle 
\psi 
 | 
\phi
\rangle
$, 
which indeed amounts to (\ref{eq:antiproduct}). 

Electron being a spin-$1/2$ particle leads one to define time reversal $\theta$ for a single-electron wave function as per $\theta = -i \sigma^y \mathcal{C}$, where  the Pauli matrix $\sigma^y$ acts on the electron spinor, and $\mathcal{C}$ is complex conjugation. This implies $\theta^2 = -1$. If $\theta$  is a symmetry and $ |\psi \rangle$ is an eigenstate, then $ |\psi \rangle$ and $ \theta |\psi \rangle$ are degenerate. But are they linearly independent? For $\mathcal{O}=1$, the equality $\theta^2 = -1$ makes the Eqn.~(\ref{eq:antiproduct2}) read  
$\langle 
\psi 
 | \theta | \psi \rangle 
 =
 - \langle
\psi 
 |
 \theta 
 |
\psi
\rangle = 0$, demonstrating orthogonality of the $ |\psi \rangle$ and $ \theta |\psi \rangle$ and thus proving the Kramers theorem: in a time reversal-invariant system, every single-electron level is at least doubly degenerate. 

The Kramers theorem above relies on the $\theta^2 = -1$ property of spin-$1/2$ particles. By contrast, in a N\'eel antiferromagnet, some of the relevant anti-unitary symmetry operators such as $\theta {\bf T_a}$, $\theta {\bf T_a U_n}(\pi)$ and $\theta {\bf T_a U_n}(\pi)\mathcal{R}_{\bf n} (\pi)$ do {\em not} square to a $C$-number -- but, rather, to an operator: for instance $[ \theta {\bf T_a} ]^2 = - {\bf T}_{2 {\bf a}}$. However, the Bloch states in question are eigenstates of such squared operators, with the eigenvalue different from unity. This guarantees degeneracy, as mentioned immediately below the Eq.~(\ref{eq:antiproduct2}).


\begin{thebibliography}{99}
\bibitem{footnote1}
The fact that the state $\theta | {\bf p} \rangle$  resides at momentum $- {\bf p}$ 
can be demonstrated by acting on $ | {\bf p} \rangle $ by an arbitrary lattice translation 
${\bf T_b}$: 
$$
{\bf T_b} \theta | {\bf p} \rangle = \theta  {\bf T_b} | {\bf p} \rangle = 
\theta  \exp \left[ i {\bf p \cdot b }\right] | {\bf p} \rangle = 
 \exp \left[ - i {\bf p \cdot b }\right]  \theta | {\bf p} \rangle .
$$
Here the first equality used the commutativity of ${\bf T_b}$ and $\theta$, 
the second used the Bloch theorem 
(${\bf T_b} | {\bf k} \rangle = \exp \left[ i {\bf k \cdot b }\right] | {\bf k} \rangle$), 
and the third one used the anti-unitarity of $\theta$. 
The orthogonality relation $\langle {\bf p} | \theta | {\bf p} \rangle = 0$ follows from the Eq. (\ref{eq:antiproduct2}) of the Appendix with $\mathcal{O} = 1$ -- and from the equality $\theta^2 = -1$, stemming from the  spin-$1/2$ nature of the electron.
\bibitem{footnote2}
The fact that the state $\mathcal{I} \theta | {\bf p} \rangle$ has momentum label ${\bf p}$ can be verified by acting on $\mathcal{I} \theta | {\bf p} \rangle$ by a translation symmetry operator ${\bf T_b}$: $${\bf T_b} \mathcal{I} \theta | {\bf p} \rangle = \mathcal{I} {\bf T}_{ -{\bf b}} \theta | {\bf p} \rangle = \mathcal{I} \theta {\bf T}_{ -{\bf b}}  | {\bf p} \rangle = \mathcal{I} \theta e^{- i {\bf b \cdot p}}  | {\bf p} \rangle = e^{ i {\bf b \cdot p}} \mathcal{I} \theta  | {\bf p} \rangle , 
$$ whereas the orthogonality relation $\langle {\bf p} | \mathcal{I} \theta  | {\bf p} \rangle = 0$ follows from the Eq. (\ref{eq:antiproduct2}) of the Appendix and from the equality $\left[ \mathcal{I \theta} \right]^2 = -1$. Notice that this result depends neither on the realization of the system nor on the choice of its eigenstates.
\bibitem{footnote.Itheta}
Notice that the four states of the quartet are mutually orthogonal only when the momenta ${\bf p}$ and $-{\bf p}$ are different in the Brillouin zone. At ${\bf p} = 0$ and ${\bf p} = \pm \pi$, only two of the four states remain linearly independent, as illustrated in the rightmost panel of the Fig. \ref{fig:PT}
\bibitem{footnote3}
The momentum labels of the three states $\mathcal{I} | {\bf p} \rangle$, 
$\theta {\bf T_a} | {\bf p} \rangle$ and $\mathcal{I} \theta {\bf T_a} | {\bf p} \rangle$  
can be found as it was done for a non-magnetic crystal, by acting on the state 
with a translation operator ${\bf T_b}$, that is a symmetry of the N\'eel state. 
As an example, consider the state 
$\mathcal{I} \theta {\bf T_a} | {\bf p} \rangle$, the others can be treated analogously: 
$$
{\bf T_b} \mathcal{I} \theta {\bf T_a} | {\bf p} \rangle  = 
\mathcal{I} \theta {\bf T_a} {\bf T}_{-{\bf b}} | {\bf p} \rangle = 
\mathcal{I} \theta {\bf T_a} \exp \left[ - i {\bf p \cdot b} \right] | {\bf p} \rangle = 
\exp \left[ i {\bf p \cdot b} \right] \mathcal{I} \theta {\bf T_a} | {\bf p} \rangle . 
$$ 
The orthogonality of the states 
$\theta {\bf T_a} | {\bf p} \rangle$ and $\mathcal{I} \theta {\bf T_a} | {\bf p} \rangle$  to 
the state $| {\bf p} \rangle$ follows from the Eq. (\ref{eq:antiproduct2}) of the Appendix, and from the equalities $\left[ \mathcal{I} \theta {\bf T_a} \right]^2 = -1$ and 
$\left[ \theta {\bf T_a} \right]^2 = - {\bf T}_{2 {\bf a}} = - \exp \left[ 2i {\bf p \cdot a} \right]$. 
The last equality holds for the action on $| {\bf p} \rangle$, since ${\bf T}_{2 {\bf a}}$ is a 
translation symmetry of the N\'eel state. The orthogonality of 
$\mathcal{I} | {\bf p} \rangle$ and $| {\bf p} \rangle$ can be demonstrated as per 
$\langle {\bf p} | \mathcal{I} | {\bf p} \rangle = 
\langle {\bf p} | \mathcal{I} {\bf T}_{-{\bf a}} {\bf T_a} | {\bf p} \rangle = 
\langle {\bf p} | {\bf T_a} \mathcal{I} {\bf T_a} | {\bf p} \rangle = 
\exp \left[ 2 i {\bf p \cdot a} \right]  \langle {\bf p} | \mathcal{I} | {\bf p} \rangle = 0$. 

As in a non-magnetic crystal, the four states of the quartet are mutually orthogonal only when the momenta ${\bf p}$ and $-{\bf p}$ are different in the Brillouin zone. At ${\bf p} = 0$ and ${\bf p} = \pm \pi$, only two of these four states are linearly independent, as illustrated in the left panel of Fig. \ref{fig:IthetaT}
\bibitem{footnote:[001]} 
Notice that for a [001] well a similar argument induces no constraint: a $\pi/2$ rotation around the growth axis [001] transforms $(J_x , J_y , J_z)$ and $(H_x , H_y , H_z)$ into $(-J_y , J_x , J_z)$ and $(-H_y , H_x , H_z)$, thus leaving the second term in the Eq. (\ref{eq:Luttinger-Zeeman}) invariant.
\end{thebibliography}
\end{document}